\documentclass[pt, letterpaper]{article}
\usepackage{amsmath}
\usepackage{amssymb}
\usepackage{graphicx}
\usepackage{placeins}
\usepackage{tikz}
\usepackage{enumitem}
\usepackage[bottom]{footmisc}

\usepackage[margin=.9in]{geometry}
\begin{document}

\begin{center}
\begin{Large}
\textbf{Social media usage patterns during natural hazards}\\[0.5\baselineskip]
\end{Large}
Meredith T. Niles$^{1,2}$, Benjamin F. Emery$^{3}$, Andrew J. Reagan$^{4}$, Peter S. Dodds$^{3}$, Christopher M. Danforth$^{3}$\\[1\baselineskip]
\end{center}
\begin{small}
$^{1}$Department of Nutrition and Food Sciences, University of Vermont\\
$^{2}$Food Systems Program, University of Vermont\\
$^{3}$Department of Mathematics \& Statistics, Vermont Complex Systems Center, University of Vermont\\
$^{4}$MassMutual Data Science, Amherst, MA 01002\\
\end{small}
\begin{changemargin}{.3in}{.3in}
\begin{small}
Natural hazards are becoming increasingly expensive as climate change and development are exposing communities to greater risks. Preparation and recovery are critical for climate change resilience, and social media are being used more and more to communicate before, during, and after disasters.  While there is a growing body of research aimed at understanding how people use social media surrounding disaster events, most existing work has focused on a single disaster case study. In the present study, we analyze five of the costliest disasters in the last decade in the United States (Hurricanes Irene and Sandy, two sets of tornado outbreaks, and flooding in Louisiana) through the lens of Twitter. In particular, we explore the frequency of both generic and specific food-security related terms, and quantify the relationship between network size and Twitter activity during disasters.  We find differences in tweet volume for keywords depending on disaster type, with people using Twitter more frequently in preparation for Hurricanes, and for real-time or recovery information for tornado and flooding events.  Further, we find that people share a host of general disaster and specific preparation and recovery terms during these events.  Finally, we find that among all account types, individuals with ``average" sized networks are most likely to share information during these disasters, and in most cases, do so more frequently than normal.  This suggests that around disasters, an ideal form of social contagion is being engaged in which average people rather than outsized influentials are key to communication. These results provide important context for the type of disaster information and target audiences that may be most useful for disaster communication during varying extreme events. \\
\end{small}
\end{changemargin}
\textbf{\textsc{I. INTRODUCTION}}\\[.5\baselineskip]
\textbf{A. Background}\\
As of 2017, it is estimated that 77\% of Americans own and use smartphones (Pew Research Center, 2018).  The adoption of this technology has given unprecedented and immediate access to people for rapidly consuming and producing information.  Commensurate with the rise in mobile communication has been a corresponding increase in the use of social media as a tool for sharing news and networking.  Eighty percent of social media use occurs via mobile technologies, and 24\% of Americans, roughly 68 million people, use Twitter (Smith et al., 2018).  Indeed, social media is increasingly changing the way society communicates before, during and after disaster events (Bagrow, 2017; Bagrow et al., 2011).  As the cost of disasters in the United States and globally continues to increase (Smith, 2018), and future climate projections indicate that extreme events will likely become more frequent and severe, disaster preparation and recovery via communication has become a critical point of study for climate change adaptation (IPCC, 2012).\\[1\baselineskip]
The use of social media for disaster communication dates at least to the Haitian earthquake of 2010, during which social media kept people around the world informed (Keim, 2011).  Evidence also suggests that the Haitian earthquake catalyzed new mechanisms of communicating about disasters, including information dissemination and crowd funding via social media (Gurman and Ellenberger, 2015; Yates and Paquette, 2011).  Since then, there has been a growing and very recent focus, both applied and academic, in understanding how social media is used during times of disasters and the ways that it may be leveraged for disaster preparedness and improving responses (Alexander, 2014).  Social media is now used by a variety of parties during disaster events including communities, governments, individuals, organizations, and media outlets, and for more than a dozen distinct purposes of communication (Houston et al. 2015). \\[1\baselineskip]
\textbf{B. Existing Research}\\
Existing research on social media and disasters has taken multiple approaches ranging from the qualitative to the quantitative. A small body of research has explored who retweets disaster information, what they retweet and why (Abdullah et al., 2017). Existing evidence indicates that people may be tweeting more frequently leading up to, during and after disaster events, and that most people are using social media via a smartphone, which enables delivery of other disaster information such as text message alerts (Stokes and Senkbeil, 2017).  Others have found that Twitter user typologies exist. Stakeholders use tweets to communicate in different ways, albeit the majority of which are dissemination of second-hand information, coordination of relief efforts, and memorialization of those affected (Takahashi et al., 2015).  Some have focused on the ethical implications of social media in disaster and crisis media, e.g. the potential for rumors to spread in such conditions (Alexander, 2014). Other research has explored the ways in which organizations, such as the Red Cross or others, have utilized social media in their disaster responses and recovery efforts (Briones et al., 2011; Gurman and Ellenberger, 2015).  \\[1\baselineskip]
More quantitative assessments have explored potential tools for disaster recovery and relief, with a particular focus on enabling effective responses.  Through Twitter analysis, researchers have examined how activity correlates with hurricane damage to potentially predict where and how to focus recovery efforts (Guan and Chen, 2014; Kryvasheyeu et al., 2015, 2016).  Others have used information learning tools to understand how people communicate on Twitter during known emergencies, to ideally enable authorities to rank social media content for prioritizing attention (Hodas et al., 2015). Relatedly, Bayesian approaches have been utilized to classify tweets during disasters as ``informational" versus ``conversational" for disaster relief prioritization (Truong et al., 2014).  Additionally, tweets during a flood disaster event were used to develop an algorithm that can identify victims asking for help via social media (Singh et al., 2017).  Research into the identity and role of individuals within a network indicates that a core set of actors served as information conduits during the Deepwater Horizon Spill (Sutton et al., 2013).\\[1\baselineskip]
Despite the growing research focus on social media use during crises, there remain many gaps in our understanding of the issue, particularly in comparison of media use across different kinds of disaster events.  In the United States, the majority of existing research focuses on only one event e.g. Hurricane Sandy - (Guan and Chen, 2014; Murthy and Gross, 2017; Truong et al., 2014), though much of the initial focus of this topic occurred following the Haitian earthquake- (Gurman and Ellenberger, 2015; Keim, 2011; Yates and Paquette, 2011). Other hazards have also been explored including Alabama tornadoes (Stokes and Senkbeil, 2017) and flood events (Singh et al., 2017). Importantly, natural disasters vary in predictability. For example, hurricane tracks are often predictable up to a week in advance, while tornadoes are typically only forecast with 15 minutes of lead time, and earthquakes even less. This suggests that who, when, and even why people communicate during different kinds of disaster events may vary and have important implications for future disaster preparedness, safety, and recovery.\\[1\baselineskip]
\textbf{C. Focus of Study}\\
We focus here on the use of Twitter as a means of communication before, during, and after five major disaster events within the United States within a recent five-year period. We are particularly interested in characterizing and understanding how the size of an individual?s social network relates to their activity during these events.  In addition to general disaster related topics, we also focus on the critical topics of food and food security during these acute events.  Food is often an important component of disaster preparation, and also one of the most immediate necessities (along with water and shelter) for disaster response.  It is common for people to prepare for known impending disasters by inundating grocery stores for basic staples like milk and bread.  This focus on food provides an opportunity to explore how the use of food-related terms varies across different kinds of disasters, ranging from those that are predicted to those that are not.  Furthermore, a focus on food is critical for future potential efforts for disaster preparedness and recovery, as climate change is expected to increase the number and intensity of extreme events such as hurricanes, floods, and tornado outbreaks, which will have significant impact on our global food systems (Brown et al., 2015). \\[1\baselineskip]
Several research questions guide our inquiry including: 1) To what extent do people tweet about emergency and disaster topics before, during, and after disaster events?  2) To what extent do people tweet about food and food security related issues before, during and after disaster events? 3) What is the relationship between the size of an individual?s social network and their activity during disaster events?  A series of disaster events reflecting a five-year time period reveals understanding and comparison of the ways in which tweeting during disaster events may vary by disaster type, contributing to a growing body of work exploring the role of social media in disaster preparation and recovery.\\[2\baselineskip]
\textbf{II. METHODS}\\[.5\baselineskip]
\textbf{A. Event selection and characteristics}\\
We utilized data from the National Centers for Environmental Information within the National Oceanic and Atmospheric Administration (NOAA), which categorizes the economic costs of weather and climate disasters (National Oceanic and Atmospheric Administration, 2016).  We focused on the most recent five years at the time of analysis, beginning in September 2016 (2011-2016) to determine the top five most costly events (consumer price index (CPI) adjusted).  Given that we are using Twitter to analyze finite disasters over short-periods of time, we excluded long-term droughts, which in this case included the U.S. drought/heatwave of 2012 (classified as lasting the entire year), the Southern Plains/Southwest drought and heatwave (Spring-Summer 2011), and the Western Plains drought/heatwave (Spring-Fall 2013). The five disasters of focus (in order of cost impacts) include Hurricane Sandy, Hurricane Irene, Southeast/ Ohio Valley/ Midwest tornadoes, Louisiana flooding, and Midwest/ Southeast tornadoes (Table 1). \\[1\baselineskip]
\textbf{B. Twitter word analysis}\\
We generated a total of 39 keyword search terms of inquiry through multiple iterations, with the aim of exploring how word usage varied on Twitter during the five disasters (Table 2).  We chose these words to represent a variety of potential tweets including: 1) food-related preparations (e.g. ``canned", ``grocery store"); 2) food-related responses (e.g. ``food pantry", ``food bank"); 3) general terms related to disaster responses (``supplies", ``emergency"); and 4) terms specific to individual disasters (e.g. ``Irene", ``Sandy", ``tornado", ``flood").\\[1\baselineskip]
\textbf{C. Keyword time-series}\\
The tweets analyzed in the present study are drawn from the version of Twitter?s streaming API commonly referred to as the `Decahose?, consisting of a random 10\% of all public messages. For each disaster event, we identified related messages authored during a two week period centered on the event, using the keywords outlined in Table 2. The frequency of each keyword was then visualized at intervals of one hour, three hours, twelve hours, and one day. Using this variable-resolution time-series data, we generated a plot of frequency vs time to visualize the changes in volume of different types of disaster-related content on Twitter during a crisis.\\[1\baselineskip]
\textbf{D. Distribution of network sizes and relation to tweet volume}\\
We examined statistics associated with the follower network of individuals who authored tweets in the collection described above. Specifically, for each tweet we use the author?s user ID, as well as the number of accounts that followed the author. Individuals with multiple messages in the two-week window were assigned the follower count associated with their first tweet. The number of messages posted by each user during the interval of interest was aggregated by user as well, representing roughly 10\% of their total number of posts. In Figure 7, we plot the base-10 logarithm of user count for a matrix of binned message frequencies and follower counts. Using account data accumulated for all disasters, we plot the total number of tweets per account against the number of followers of the account using both linear and logarithmic scales. While the follower count is not a proxy for meaningful interaction, it is a first order approximation of the size of an account?s audience.\\[1\baselineskip]
\textbf{E. Tweet volume increase by network size}\\
We estimated changes in individual behavior observed during Hurricane Sandy, compared to a baseline reference, as a function of network size. To do this, we used the ``total tweet count" field in the Decahose JSON metadata, which represents the exact number of messages posted to the account up to that moment. For each user found to have tweeted one of the keywords surrounding Hurricane Sandy?s landfall, we collected the first and last tweet authored during the month of September 2012. We used these two tweets to compute a baseline tweet rate, found by taking the difference in total tweet count and dividing by the number of days between the two tweets. We repeated this process for October 21 through November 4, 2012 to compute the tweet rate during the disaster and its aftermath. We required at least two tweets during each period for a user to be included. We used the quotient of the tweet rate during Sandy (numerator) and the baseline tweet rate (denominator) to compute the estimated change in tweet volume for each user. Despite being restricted to a random 10\% of messages, and therefore not being able to observe most tweets, the message rate calculation is exact for the period of observation. Work by Barabasi has shown that the rates of human activities such as emailing follow a Pareto distribution of lag time between events (2005). Although tweeting likely follows a similar distribution, our sample does not allow us to accurately measure the lag time between user tweets, so we are limited to this homogeneous approach. 
\newpage
\begin{center}
\begin{small}
Table 1. Weather and Climate Billion-Dollar Hazard Events to affect the U.S. from 2011-2016 (CPI-Adjusted)\\
\begin{tabular}{|c|c|l|c|c|}
\hline
\textbf{Event}                                                                            & \textbf{Dates}                                                      & \multicolumn{1}{c|}{\textbf{Direct Summary (from NOAA)}}                                                                                                                                                                                                                                                                                                                                                                                                                                                                                                                                                                                                                                                          & \textbf{\begin{tabular}[c]{@{}c@{}}CPI-\\ Adjusted \\ Estimated \\ Cost (in \\ Billions)\end{tabular}} & \textbf{Deaths} \\ \hline
\begin{tabular}[c]{@{}c@{}}Hurricane\\ Sandy\end{tabular}                                 & \begin{tabular}[c]{@{}c@{}}10/30/2012\\ -\\ 10/31/2012\end{tabular} & \begin{tabular}[c]{@{}l@{}}"Extensive damage across several northeastern states\\ (MD, DE, NJ, NY, CT, MA, RI) due to high wind \\ and coastal storm surge, particularly NY and NJ.\\ Damage from wind, rain and heavy snow also \\ extended more broadly to other states (NC, VA, WV, \\ OH, PA, NH), as Sandy merged with a developing \\ Nor'easter. Sandy's impact on major population \\ centers caused widespread interruption to critical \\ water / electrical services and also caused 159 deaths \\ (72 direct, 87 indirect). Sandy also caused the New \\ York Stock Exchange to close for two consecutive \\ business days, which last happened in 1888 due to a \\ major winter storm."\end{tabular} & \$70.9 CI                                                                                              & 159             \\ \hline
\begin{tabular}[c]{@{}c@{}}Hurricane\\ Irene\end{tabular}                                 & \begin{tabular}[c]{@{}c@{}}8/26/2011\\ -\\ 8/28/2011\end{tabular}   & \begin{tabular}[c]{@{}l@{}}"Category 1 hurricane made landfall over coastal NC \\ and moved northward along the Mid-Atlantic Coast \\ (NC, VA, MD, NJ, NY, CT, RI, MA, VT) causing \\ torrential rainfall and flooding across the Northeast. \\ Wind,damage in coastal NC, VA, and MD was \\ moderate with considerable damage resulting from \\ falling trees and power lines, while flooding caused \\ extensive flood damage across NJ, NY, and VT. Over \\ seven million homes and businesses lost power during \\ the storm. Numerous tornadoes were also reported in \\ several states further adding to the damage."\end{tabular}                                                                          & \$15.1 CI                                                                                              & 45              \\ \hline
\begin{tabular}[c]{@{}c@{}}Southeast/\\ Ohio\\ Valley/\\ Midwest\\ Tornadoes\end{tabular} & \begin{tabular}[c]{@{}c@{}}4/25/2011\\  -\\ 4/28/2011\end{tabular}  & \begin{tabular}[c]{@{}l@{}}"Outbreak of tornadoes over central and southern \\ states (AL, AR, LA, MS, GA, TN, VA, KY, IL, MO, \\ OH, TX,OK) with an estimated 343 tornadoes. \\ The deadliest tornado of the outbreak, an EF-5, hit \\ northern Alabama, killing 78 people. Several major \\ metropolitan areas were directly impacted by strong \\ tornadoes including Tuscaloosa, Birmingham, and \\ Huntsville in Alabama and Chattanooga, Tennessee, \\ causing the estimated damage costs to soar."\end{tabular}                                                                                                                                                                                            & \$11.4 CI                                                                                              & 321             \\ \hline
\begin{tabular}[c]{@{}c@{}}Louisiana\\ Flooding\end{tabular}                              & \begin{tabular}[c]{@{}c@{}}8/12/2016\\ -\\ 8/15/2016\end{tabular}   & \begin{tabular}[c]{@{}l@{}}"A historic flood devastated a large area of southern \\ Louisiana resulting from 20 to 30 inches of rainfall \\ over several days. Watson, Louisiana received an \\ astounding 31.39 inches of rain from the storm. Two-\\ day rainfall totals in the hardest hit areas have a 0.2\% \\ chance of occurring in any given year: a 1 in 500 year \\ event. More than 30,000 people were rescued from\\ the floodwaters that damaged or destroyed over \\ 50,000 homes, 100,000 vehicles and 20,000 \\ businesses. This is the most damaging U.S. flood \\ event since Superstorm Sandy impacted the Northeast \\ in 2012."\end{tabular}                                                 & \$10.4 CI                                                                                              & 13              \\ \hline
\begin{tabular}[c]{@{}c@{}}Midwest/\\ Southeast\\ Tornadoes\end{tabular}                  & \begin{tabular}[c]{@{}c@{}}5/22/2011\\ -\\ 5/27/2011\end{tabular}   & \begin{tabular}[c]{@{}l@{}}"Outbreak of tornadoes over central and southern \\ states (MO, TX, OK, KS, AR, GA, TN, VA, KY, IN, \\ IL, OH, WI, MN, PA) with an estimated 180 \\ tornadoes. Notably, an EF-5 tornado struck Joplin, \\ MO resulting in at least 160 deaths, making it the \\ deadliest single tornado to strike the U.S. since \\ modern tornado record keeping began in 1950."\end{tabular}                                                                                                                                                                                                                                                                                                        & \$10.2 CI                                                                                              & 177             \\ \hline
\end{tabular}
\end{small}
\end{center}
\newpage
\begin{center}
\begin{small}
Table 2. Words used in the Twitter analysis\\
\begin{tabular}{|l|l|l|l|l|l|l|l|}
\hline
canned    & food assistance & food security & fridge        & hurricane & rain    & snow        & unprepared \\ \hline
drinks    & food bank       & food shelf    & generator     & irene     & sandy   & store       & water      \\ \hline
emergency & food insecurity & food stamps   & groceries     & power     & shelter & supermarket & watson     \\ \hline
farm      & food market     & food store    & grocery store & prepare   & shock   & supplies    & wind       \\ \hline
flood     & food pantry     & foods         & help          & preparing & SNAP    & tornado     &            \\ \hline
\end{tabular}
\end{small}
\end{center}
\noindent We also generated a null-model of this change in tweet volume by using the same method to estimate the change for the same users between every month of the year and the following 16-day period, analogous to the dates we sampled for Hurricane Sandy. These pairs of time periods were all observed in 2012, except for those that overlapped with Hurricane Sandy, which were instead drawn from 2011.\\[1\baselineskip]
\textbf{F. Temporal tweet distribution analysis}\\
We conducted a content analysis to describe word use during the five disasters. We analyzed Tweet count distributions for 39 words across the five disaster events, and word use was categorized by each disaster if it appeared with above average activity compared to the baseline.  For the distributional analysis we omitted four words (`drinks?, `snow?, `store?, `watson?) for analysis, as there was no discernible difference in tweet volume across any disaster or timescale.  We draw upon Murthy and Gross (2017) who explored the evolution of disaster tweets in the lead up to the event (anticipatory), the core event, and the aftermath (Figure 1) (Murthy and Gross, 2017).  Using this framework, we determined five categories for word frequency based on above average use during the disaster: 1) before the disaster (anticipatory); 2) before and during the disaster (anticipatory and core event); 3) during the disaster (core event); 4) during and after the disaster (core event and aftermath); and 5) after the disaster (aftermath).\\
\begin{center}
\includegraphics[width=\textwidth]{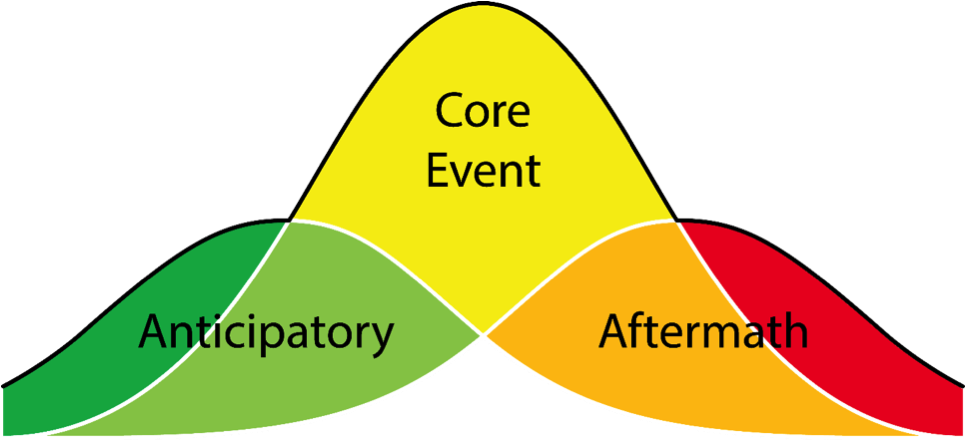}
\end{center}
\begin{small}
Figure 1.  Framework for assessing tweets in the context of disasters.  Adapted from Murthy and Gross 2017.\\[2\baselineskip]
\end{small}
\textbf{III. RESULTS}\\[.5\baselineskip]
\textbf{A. Tweet frequencies}\\
We analyze tweet frequencies for our chosen 39 words over the two-week period of the event (before, during and after).  Given the volume of data (39 words, across five disasters, measured at four timescale intervals), the entire dataset is contained in the Supplementary Materials.  Here we center on a few key results.\\[1\baselineskip]
We find that tweets containing certain words occur across all of the disasters that we study, albeit with varying frequency and across different timescales throughout the two week period.  General terms such as emergency, flood, hurricane, shelter, tornado, water and wind are used frequently before, during, or after the events (Figure 2 shows example plots for ``emergency").  Consistent across all of these general terms related to disasters or specific kinds of weather and impacts, we find they are most frequently tweeted before Hurricane Sandy occurs, as opposed to during or after the event as was the case with the other disasters.  \\[1\baselineskip]
\includegraphics[width=\textwidth]{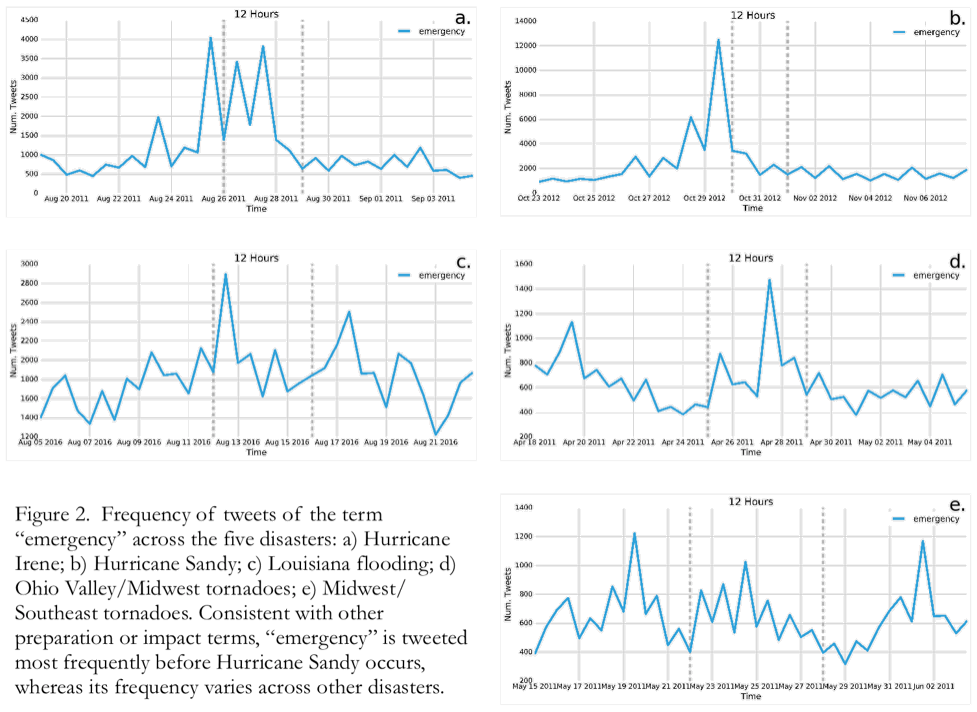}\\[1\baselineskip]
Our examination of more specific terms related to preparation or event impacts (e.g. ``generator", ``power", ``prepare") yields similar results, namely that the ways these terms are used on Twitter depends on the event.  Similar to general terms examined above, we also find consistent evidence that tweets using these terms during Hurricane Sandy were used most frequently before the event began, indicating conversation and sharing about preparation (Figure 3).  Conversely, it was more  common that these terms were used during Hurricane Irene and during or after the Louisiana flooding.  In the case of the tornado events, in some cases these terms showed no clear increase in their use (e.g. ``generator"). \\[1\baselineskip] 
\includegraphics[width=\textwidth]{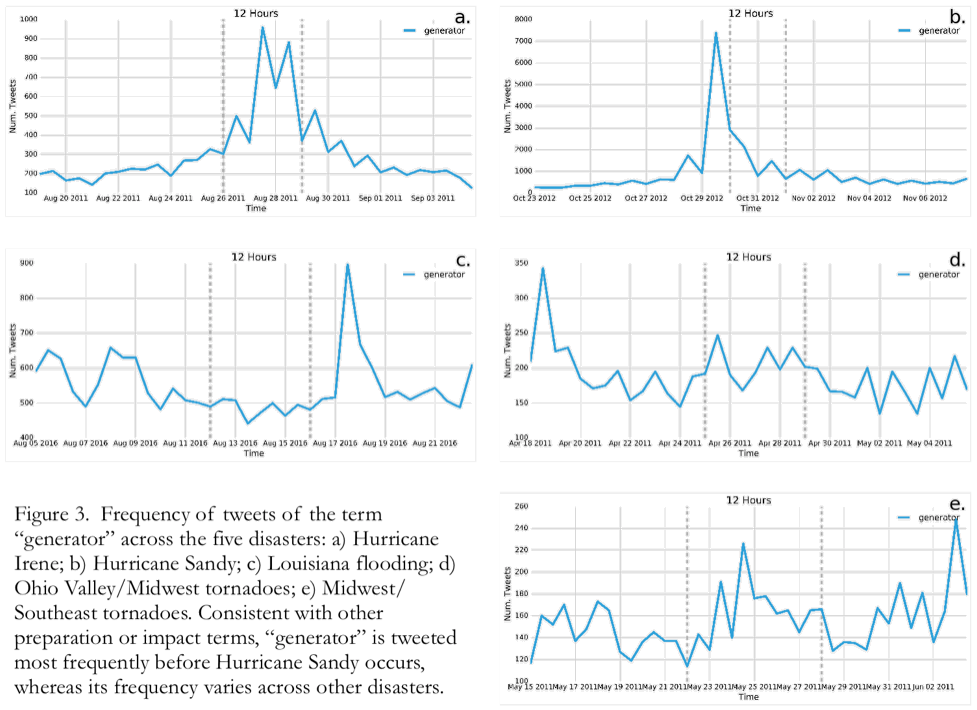}\\[2\baselineskip]
As it relates to food specific terms, we highlight here three food related terms that indicate how people discuss food security in the context of acute disaster events via Twitter.  In terms of preparation, we find that the use of ``supermarket" occurs in events that were anticipated, including Hurricane Irene and Hurricane Sandy (Figure 4).  Notably, the increase in ``supermarket" tweets occurred as Hurricane Irene was happening, whereas it occurred two to three days before Hurricane Sandy occurred.  We do not find a notable increase in the use of the term ``supermarket" in the Louisiana flooding though we do see a potential increase in tweets compared to baseline in the instances of the tornadoes.  We also find evidence of some food related terms used more frequently after the events. ``Food bank" was consistently used in greater quantities than baseline in all of the events, either during or after the event, suggesting its use as a means of communicating about food availability for those impacted by the disasters.  We also tested other phrases synonymous with ``food bank", such as ``food shelf", which at least in some cases demonstrated similar results.  The use of ``food stamps" occurs notably in the case of Hurricane Irene.\footnote{Instances of ``food stamps'' were most notable nearly one week after Hurricane Sandy.  However, since this was also a presidential election day, we are uncertain that this was related to Hurricane Sandy.} \\[1\baselineskip]
\includegraphics[width=\textwidth]{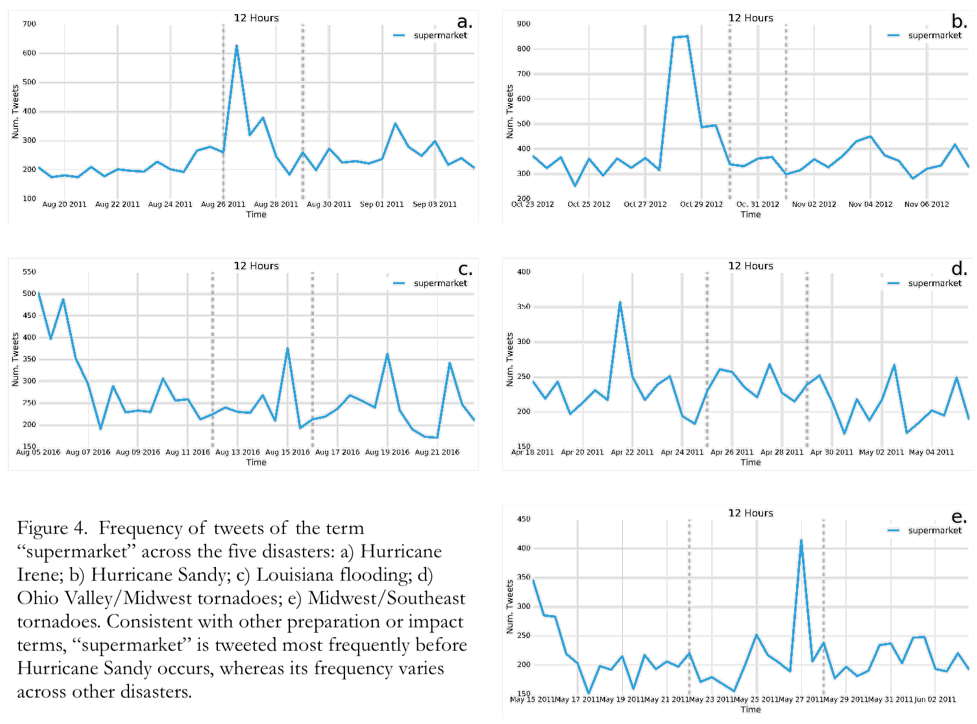}\\[2\baselineskip]
\textbf{B. Temporal tweet distribution}\\
In addition to exploring the words that were used before, during, and after tweets, we also were interested in exploring whether certain events had increased tweets before, during, or after events.  We expected that events that were more predictable (e.g., hurricanes) would be more likely to have greater tweet counts than average before events, whereas events like tornadoes would be more likely to have greater tweet volume during or after an event.  Figure 5 demonstrates the analysis of five disasters and our total word count across these temporal distributions, documenting cases in which there was a notable increase in tweet volume as compared to baseline.  We find evidence for our hypothesis in that tweet volume was much higher before Hurricane Sandy, and higher before and during Hurricane Irene.  During Hurricane Sandy 76\% (19/25) of searched words peaked as anticipatory or anticipatory/core event tweets.  For Irene, 50\% (16/32) peaked as anticipatory or anticipatory/core event tweets.  Conversely, we find that tweets during the tornado events were collectively most likely to occur during the core event or core event/aftermath.  In the first round of tornadoes, 72\% (18/25) peak word tweets occurred as core event or core event/aftermath tweets.  We find similar results for the second round of tornadoes where 64\% (16/25) keyword tweets peaked during core event or core event/aftermath.  Finally, we find that for the Louisiana flooding, which was not widely predicted or communicated, tweet volume was most likely to occur in the core event/aftermath (32\%) and aftermath (29\%) timeframe. \\[1\baselineskip]
\includegraphics[width=\textwidth]{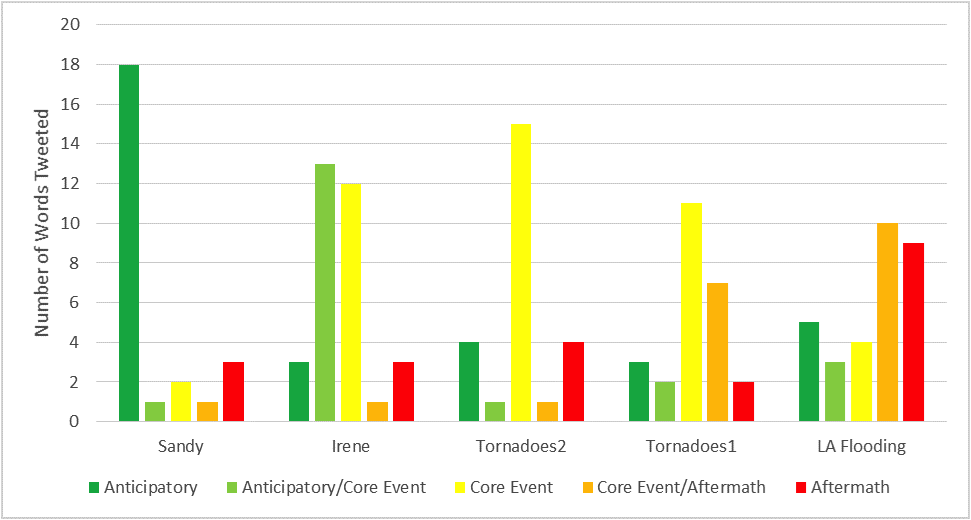}\\
\begin{small}
Figure 5. Peak distributions above baseline of keywords in tweets across the five disasters.  Analysis based on the framework in Murthy and Gross 2017. \\[2\baselineskip]
\end{small}
\textbf{C. Twitter networks}\\
We also next sought to understand the relationship between tweet frequency and follower count. While the follower count associated with an individual is not a perfect reflection of their influence, it does serve as a proxy for the size of their audience. In looking at the follower counts associated with individuals tweeting about disaster events, we are seeking an understanding of the role various stakeholders play in the spread of information.\\[1\baselineskip]
\includegraphics[width=\textwidth]{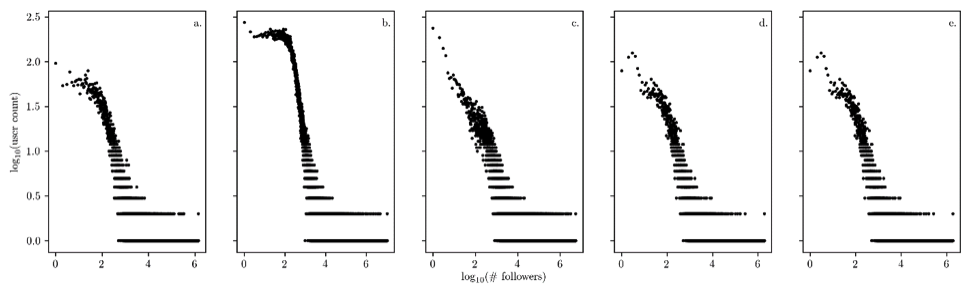}\\
\begin{small}
Figure 6. Log-log plot of the distribution of individuals observed to tweet during each disaster as a function of their follower count across the five disasters: a) Hurricane Irene; b) Hurricane Sandy; c) Louisiana flooding; d) Ohio Valley/Midwest tornadoes; e) Midwest/Southeast tornadoes. Most accounts have a small number of followers (e.g. less than 100), and a few accounts have many followers (e.g. more than 10,000).\\[1\baselineskip]
\end{small}
In Figure 6, we plot the distribution of follower counts, which appear typical for social networks. To explore user behavior further, we establish a baseline tweet rate for each account, and observe the increase (or decrease) in activity during the disaster event. We find a consistent trend that the individuals who tweet the most during disaster events tend to have ``average" sized networks  (Figure 7). Goncalves et al. (2011) found that social networks reflect Dunbar?s number, leading an individual?s set of meaningful relationships to be limited to between 100 and 200 accounts. It is these accounts in which we see the largest increase in activity during disasters (Gonçalves et al., 2011). Previous work has found that ``hidden influentials" in social networks, which are users with average-sized audiences, are key to allowing system-wide information-cascades and therefore play a major role facilitating protests online (González-Bailón et al. 2013, Baños et al., 2013, Borge-Holthoefer et al., 2016).\\[1\baselineskip]
Our analysis also suggests that individuals were tweeting more frequently during Sandy than during other disaster events that we studied.  Further analysis (Figures 7 and 8) explores how the tweet rate changed as compared to baseline during Hurricane Sandy.  In Figure 7 we see that, while the distribution of tweet rate change between two time periods is normally symmetric about 0 for users of all follower counts, this distribution for Hurricane Sandy is shifted upwards for users with 100 followers or fewer. In Figure 8, the same tweet rate increase distribution is shown, but the follower counts of the users are discretized by order of magnitude (0-10,10-100,100-1000,?). This demonstrates that while users of all follower counts tend to have no change in tweet rate during a typical baseline period (right panel), during Hurricane Sandy a positive average tweet rate change is observed for all users (left panel). More notably, the average tweet rate change is, slightly but significantly, higher during Sandy for users in the first and second groups: those with follower counts between 0 and 100. These results suggest that people with average-sized networks were more likely to tweet with a higher relative frequency during Hurricane Sandy than those with larger networks. \\[1\baselineskip]
\includegraphics[width=\textwidth]{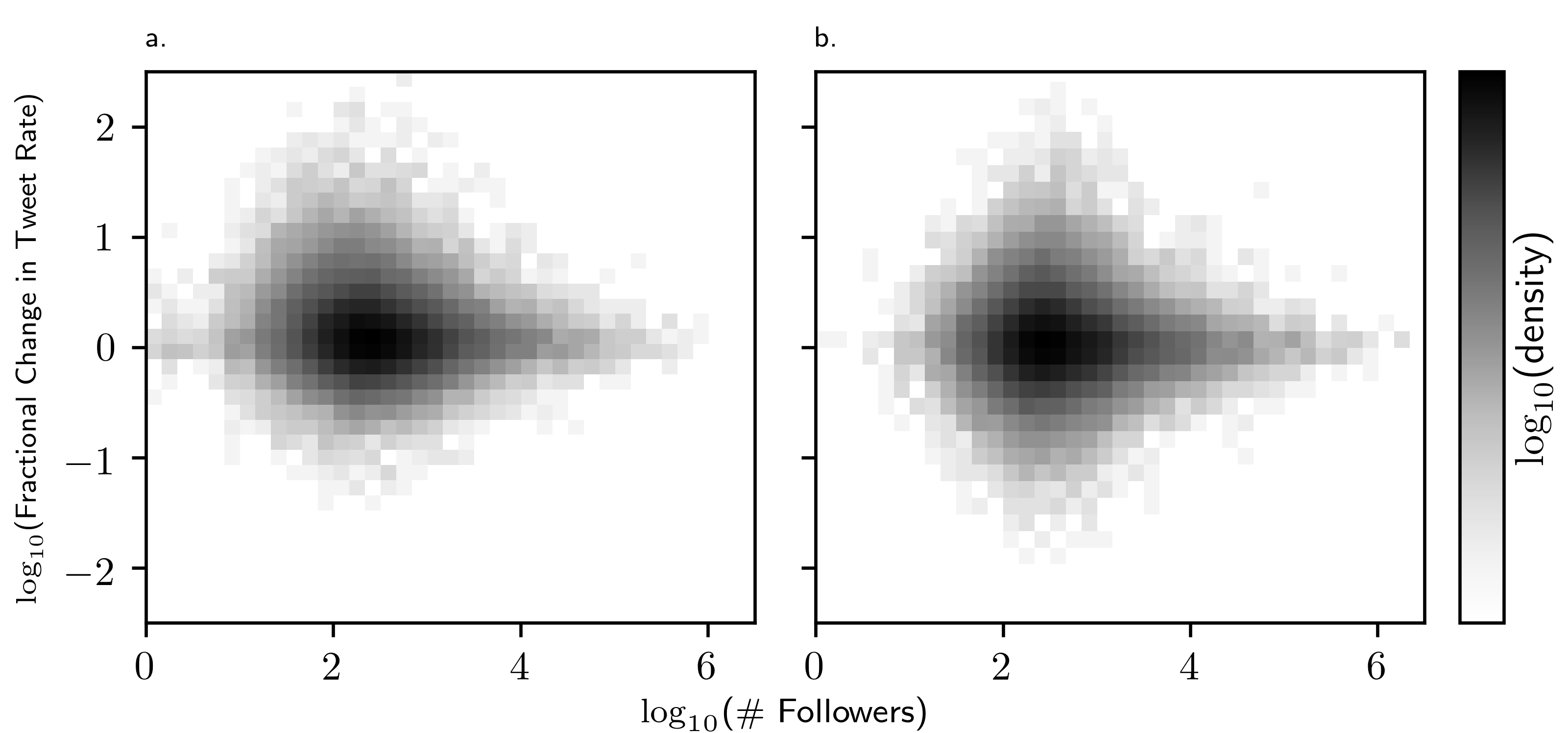}\\[1\baselineskip]
\begin{small}
Figure 7. Log-log plot of the fractional change in tweet rate as a function of follower count for (a) before and during Hurricane Sandy and (b) the pairs of times collected for the null distribution. The increased density observed above 0 suggests that most individuals tweet more frequently during the disaster. In addition, the rate increase is largest for ``average" individuals, i.e. those with 100 followers or fewer. . This is of notable contrast to the null distribution, which is roughly symmetric about the zero-axis. Note that white pixels indicate one or zero individuals exhibiting the corresponding rate change. \\[1\baselineskip]
\end{small}
\includegraphics[width=\textwidth]{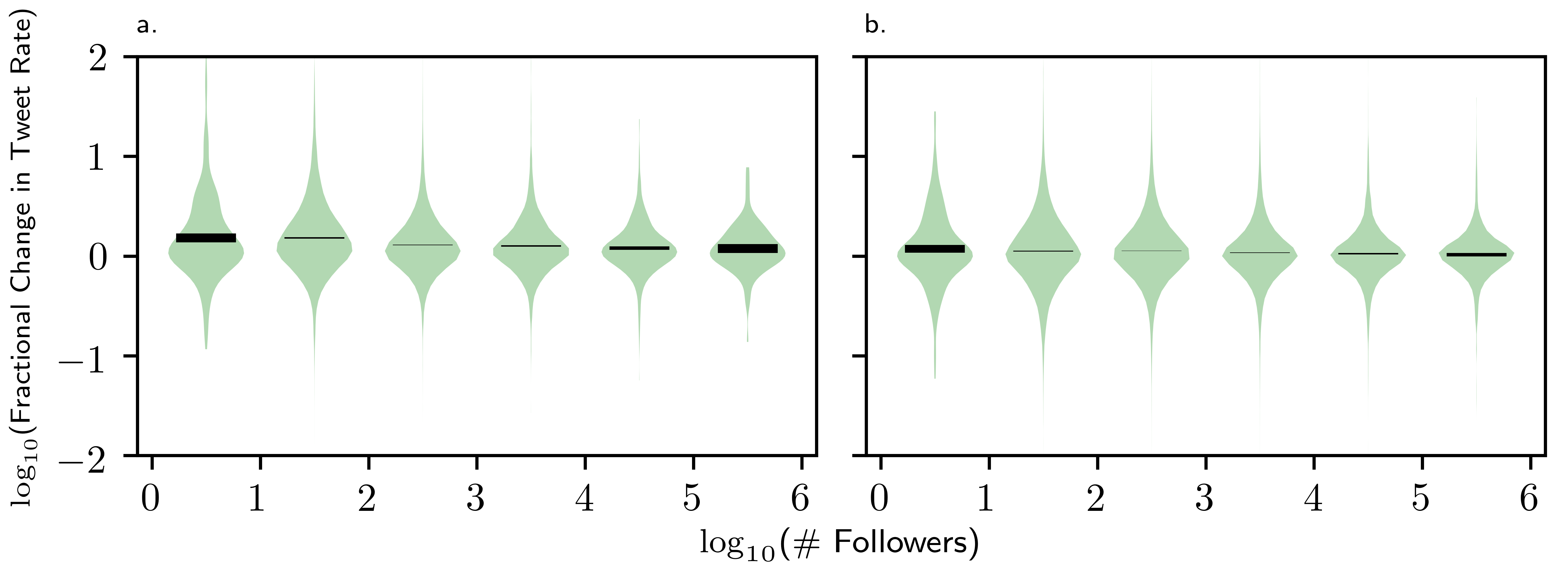}\\[1\baselineskip]
\begin{small}
Figure 8. Violin plots showing the distributions of fractional tweet rate change of the users found to be tweeting about Hurricane Sandy as it occurred (a) before and during Hurricane Sandy and (b) the pairs of times collected for the null distribution. Separate violin diagrams are drawn for users whose follower counts fall into each order of magnitude from $10^{0}$ to $10^{5}$. On each violin, a black bar indicates a Bayesian 95\% confidence interval for the mean of the population distribution given the sample. For the Hurricane Sandy data, the intervals for $10^{0}$ and $10^{1}$ are both notably higher than, and don?t overlap with the intervals for any of the higher orders of magnitude in follower count. The same is not true for the null distributions, for which most of the intervals overlap and are generally closer together. The values of endpoints of the intervals are given in the supplementary materials.\\[2\baselineskip]
\end{small}
\textbf{IV. DISCUSSION}\\[1\baselineskip]
This work explores the extent and content of tweets across five different disasters representing three types of disaster events during a recent five-year period.  We find that there are a variety of terms that individuals tweet about related to disaster events including general disaster terms, specific terms related to preparation for disasters and recovery efforts.  Across these kinds of events we find many references to specific food-related terms that people use to discuss both preparation and recovery efforts, demonstrating that Twitter is being used to discuss or provide information about food in both the preparation for and recovery from many kinds of disaster events.  As such, Twitter can be used to provide information about food recovery events and proper planning before a disaster is imminent, if time allows.\\[1\baselineskip]
Importantly, we compare Twitter activity across five disaster events, providing insight into the ways that Twitter activity varies by disaster type.  We find that when people tweet is related to the nature of the disaster- disasters that are more predictable see tweets occurring before and during, suggesting tweets of preparation.  This is especially true of Hurricane Sandy, which was predicted well in advance.  Interestingly, we find that it is the ``average" Twitter user that are tweeting most frequently during disaster events, which would suggest that this core set of actors (Sutton et al. 2013, Watts and Dodds 2007) disseminating knowledge, is in fact ``average" users.   However, at least in the case of Hurricane Sandy, people are tweeting more frequently during disaster events than normal.  These results are consistent with Stokes and Senkbeil (2017), who also found that individuals tweeted more frequently during tornado outbreaks.  It also suggests the complexity of Twitter use during disaster events, as Truong et al. 2014 highlight the use of Twitter both for conversational and informational purposes.\\[1\baselineskip]
We suggest that there are two key implications of this work.  First, identifying varying tweet activity based on the disaster type is important for disaster preparedness and recovery.  For disaster events that are predicted ahead of time such as hurricanes, Twitter can be used as a valuable tool for sharing preparation or evacuation information.  For events that are less predictable, such as tornadoes, it appears from our work that Twitter is being used in the immediate timeframe to communicate about the emergency and related terms.  Recognizing how people may use social media as a tool for disaster preparation, to share information in the immediate-term, and also for recovery efforts is critical for disaster communication networks.\\[1\baselineskip]
Furthermore, our finding that ``average" people tweet the most indicates that it may not be individuals or organizations with high numbers of followers that are critical for disseminating information, but rather that it is ``everyday" people who are disseminating information.  In this context, it may be the case that general tweets aimed at a broad audience about information could be most useful for disseminating through social networks, rather than targeting core actors in disaster events. This comports strongly with the general observation that the large-scale social contagion is necessarily mediated through networks by average individuals rather than so-called ``influentials" (Watts and Dodds 2007). With this framing, the communication strategy focuses on creating messages that will (1) impart key information to individuals and (2) be messages that average people will feel compelled to share. Moreover, the messages must be robust in the ``social wild" and not degrade through some version of the ``telephone game". Crucially, messages must contain easy-to-use links, phone numbers, etc., that connect people to vital central sources of trusted information.\\[1\baselineskip]
Our study has several limitations, some of which we address here. First, tweets represented a small subset of the daily communication made by a large but non-representative sample of individuals. Furthermore, our analysis only includes a random 10\% of messages, and as a result our study does not reflect an exhaustive characterization of society?s response to natural disasters on Twitter.  Second, we do not focus specifically on geolocated tweets, effectively including in our sample individuals not proximate to the events in question. While people geographically close to a natural hazard event are clearly most likely to be affected, information about the events can be shared worldwide and it is this sharing we hope to focus on. Third, we do not remove automated account activity from our sample. Twitter removes automated accounts reported to post abusive content, but the ecosystem of ``bots" is quite diverse, making it increasingly difficult to algorithmically segregate all but the most obvious malicious behavior (Clark et al. 2016). Indeed, human raters visiting the account page associated with a handle struggle to agree on whether it reflects robotic activity (Varol et al. 2017). Finally, we do not have retroactive network information associated with accounts, and are not able to verify the exact mechanism by which individuals are first exposed to information related to each disaster. Our study is observational, and no causal effects can be inferred.  Nevertheless, our results offer new insight into the variation of Twitter activity across many disaster and content types, providing important suggestions for improving disaster communication for preparation and recovery efforts.\\[1\baselineskip]
\textbf{V. CONCLUSION}\\[1\baselineskip]
Future climate predictions indicate that weather-related disasters will increase in both severity and frequency in the coming decades (IPCC 2012).  As such, social media is becoming an increasingly important tool to help people prepare for and recover from disasters. In particular, the mechanisms by which information is shared across networks during disaster events can have significant implications for disaster damages and recovery. Our work finds that the timeframes in which people communicate on Twitter varies by the kind of disaster event.  Furthermore, we find that it is people with ``average" sized Twitter networks that tweet most frequently during disaster events.  We find that people tweet about general disaster terms, specific preparations and recovery and a suite of food-related terms for both preparation and recovery.  Each of these findings provides insight into potential strategies for disaster communication, based on both the disaster context and the importance of general messaging that is applicable to typical Twitter users.  Such information can be useful for planning for future disasters and enabling effective recovery following disasters, which will ideally minimize disaster damages and help increase resilience in a changing climate.\\[1\baselineskip]
\textbf{ACKNOWLEDGEMENTS}\\[1\baselineskip]
The authors thank Serge Wiltshire at the University of Vermont for his graphic assistance in creating Figure 1.\\[1\baselineskip]
\textbf{REFERENCES}
\begin{itemize}[label={},leftmargin=15pt,labelindent=5pt,itemindent=-15pt]
\item Abdullah, N.A., Nishioka, D., Tanaka, Y., Murayama, Y., 2017. Why I Retweet? Exploring User?s Perspective on Decision-Making of Information Spreading during Disasters, in: Proceedings of the 50th Hawaii International Conference on System Sciences. p. 10.\\
\item Alexander, D.E., 2014. Social Media in Disaster Risk Reduction and Crisis Management. Sci. Eng. Ethics 20, 717-733. doi:10.1007/s11948-013-9502-z\\
\item Bagrow, J.P., 2017. Information spreading during emergencies and anomalous events. arXiv:1703.07362.\\
\item Bagrow, J.P., Wang, D., Barabási, A.-L., 2011. Collective Response of Human Populations to Large-Scale Emergencies. PLoS One 6, e17680.\\
\item Baños, R. A., Borge-Holthoefer, J., Moreno, Y., 2013. The role of hidden influentials in diffusion of online information cascades. EPJ Data Science, 2, 6. https://doi.org/10.1140/epjds18\\
\item Barabási, A. L., 2005. The origin of bursts and heavy tails in human dynamics. Nature, 435(7039),207. http://dx.doi.org/10.1038/nature03459\\
\item Borge-Holthoefer, J., Perra, N., Gonçalves, B., González-Bailón, S., Arenas, A., Moreno, Y., \& Vespignani, A. (2016). The dynamics of information-driven coordination phenomena: A transfer entropy analysis. Science advances, 2(4), e1501158.\\
\item Briones, R.L., Kuch, B., Liu, B.F., Jin, Y., 2011. Keeping up with the digital age: How the American Red Cross uses social media to build relationships. Public Relat. Rev. 37, 37-43. \\doi:https://doi.org/10.1016/j.pubrev.2010.12.006\\
\item Brown, M.E., Antle, J.M., Backlund, P., Carr, E.G., Easterling, W.E., Walsh, M.K., Ammann, C., Attavanich, W., Barrett, C.B., Bellemare, M.F., Dancheck, V., Funk, C., Grace, K., Ingram, J.S.I., Jiang, H., Maletta, H., Mata, T., Murray, A., Ngugi, M., Ojima, D., O?Neill, B., Tebaldi, C., 2015. Climate Change, Global Food Security, and the U.S. Food System (Report). USDA Technical Document, Washington DC. doi:10.7930/J0862DC7\\
\item Gonçalves, B., Perra, N., Vespignani, A., 2011. Modeling Users? Activity on Twitter Networks: Validation of Dunbar?s Number. PLoS One 6, e22656.\\
\item Gonzáles-Bailón, S. Borge-Holthoefer, J., Moreno, Y., 2013. Broadcasters and hidden influentials in online protest diffusion. American Behavioral Scientist, 57(7), 943-965. doi 10.1177/0002764213479371\\
\item Guan, X., Chen, C., 2014. Using social media data to understand and assess disasters. Nat. Hazards 74, 837-850. doi:10.1007/s11069-014-1217-1\\
\item Gurman, T.A., Ellenberger, N., 2015. Reaching the Global Community During Disasters: Findings From a Content Analysis of the Organizational Use of Twitter After the 2010 Haiti Earthquake. J. Health Commun. 20, 687-696. doi:10.1080/10810730.2015.1018566
\item Hodas, N.O., Ver Steeg, G., Harrison, J., Chikkagoudar, S., Bell, E., Corley, C.D., 2015. Disentangling the lexicons of disaster response in twitter, in: Proceedings of the 24th International Conference on World Wide Web. ACM, pp. 1201-1204.\\
\item Houston, J.B., Hawthorne, J., Perreault, M.F., Park, E.H., Goldstein Hode, M., Halliwell, M.R., Turner McGowen, S.E., Davis, R., Vaid, S., McElderry, J.A., Griffith, S.A., 2015. Social media and disasters: a functional framework for social media use in disaster planning, response, and research. Disasters 39, 1-22. doi:10.1111/disa.12092\\
\item IPCC, 2012. Summary for Policymakers, in: Field, C.B., Barros, V., Stocker, T.F., Qin, D., Dokken, D.J., Ebi, K.L., Mastrandrea, M.D., Mach, K.J., Plattner, G.-K., Allen, S.K., Tignor, M., Midgley, P.M. (Eds.), Managing the Risks of Extreme Events and Disasters to Advance Climate Change Adaptation. Cambridge University Press, Cambridge, UK; New York, NY.\\
\item Keim, M., 2011. Emergent use of Social Media: A New Age of Opportunity for Disaster Resilience. Prehosp. Disaster Med. 26, s94-s94. doi:DOI: 10.1017/S1049023X11003190\\
\item Kryvasheyeu, Y., Chen, H., Moro, E., Van Hentenryck, P., \& Cebrian, M. (2015). Performance of social network sensors during Hurricane Sandy. PLoS one, 10(2), e0117288.\\
\item Kryvasheyeu, Y., Chen, H., Obradovich, N., Moro, E., Van Hentenryck, P., Fowler, J., Cebrian, M., 2016. Rapid assessment of disaster damage using social media activity. Sci. Adv. 2.\\
\item Murthy, D., Gross, A.J., 2017. Social media processes in disasters: Implications of emergent technology use. Soc. Sci. Res. 63, 356-370. doi:https://doi.org/10.1016/j.ssresearch.2016.09.015\\
\item National Oceanic and Atmospheric Administration, 2016. Billion-dollar weather and climate disasters [WWW Document]. 2011-2016 Events. URL https://www.ncdc.noaa.gov/billions/events/US/2011-2016 (accessed 9.30.16).\\
\item Pew Research Center, 2018. Mobile Fact Sheet [WWW Document]. URL http://www.pewinternet.org/fact-sheet/mobile/ (accessed 4.23.18).\\
\item Singh, J.P., Dwivedi, Y.K., Rana, N.P., Kumar, A., Kapoor, K.K., 2017. Event classification and location prediction from tweets during disasters. Ann. Oper. Res. 1-21.\\
\item Smith, A., Anderson, M., Calazza, T., 2018. Social media use in 2018.\\
\item Smith, A.B., 2018. 2017 U.S. billion-dollar weather and climate disasters: a historic year in context. Clim. Mag.\\
\item Stokes, C., Senkbeil, J.C., 2017. Facebook and Twitter, communication and shelter, and the 2011 Tuscaloosa tornado. Disasters 41, 194-208. doi:10.1111/disa.12192\\
\item Sutton, J., Spiro, E., Butts, C., Fitzhugh, S., Johnson, B., Greczek, M., 2013. Tweeting the Spill: Online Informal Communications, Social Networks, and Conversational Microstructures during the Deepwater Horizon Oilspill. Int. J. Inf. Syst. Cris. Response Manag. 5, 58-76. doi:10.4018/jiscrm.2013010104\\
\item Takahashi, B., Tandoc, E.C., Carmichael, C., 2015. Communicating on Twitter during a disaster: An analysis of tweets during Typhoon Haiyan in the Philippines. Comput. Human Behav. 50, 392-398. \\doi:https://doi.org/10.1016/j.chb.2015.04.020\\
\item Truong, B., Caragea, C., Squicciarini, A., Tapia, A.H., 2014. Identifying valuable information from twitter during natural disasters. Proc. Am. Soc. Inf. Sci. Technol. 51, 1-4. doi:10.1002/meet.2014.14505101162\\
\item Yates, D., Paquette, S., 2011. Emergency knowledge management and social media technologies: A case study of the 2010 Haitian earthquake. Int. J. Inf. Manage. 31, 6-13. \\doi:https://doi.org/10.1016/j.ijinfomgt.2010.10.001\\
\end{itemize}

\end{document}